# Securing the Future: Proactive Threat Hunting for Sustainable IoT Ecosystems

Saied Ghasemshirazi
saeidghasemshirazi@cmail.carleton.ca
Carleton University
Ottawa, Ontario, Canada

Ghazaleh Shirvani
ghazalehshirvani@cmail.carleton.ca
Carleton University
Ottawa, Ontario, Canada

## ABSTRACT
In the rapidly evolving landscape of the IoT, the security of connected devices has become a paramount concern. This paper explores the concept of proactive threat hunting as a pivotal strategy for enhancing the security and sustainability of IoT systems. Proactive threat hunting is an alternative to traditional reactive security measures that analyses IoT networks continuously and in advance to find and eliminate threats before they occure. By improving the security posture of IoT devices this approach significantly contributes to extending IoT operational lifespan and reduces environmental impact. By integrating security metrics similar to the Common Vulnerability Scoring System (CVSS) into consumer platforms, this paper argues that proactive threat hunting can elevate user awareness about the security of IoT devices. This has the potential to impact consumer choices and encourage a security-conscious mindset in both the manufacturing and user communities. Through a comprehensive analysis, this study demonstrates how proactive threat hunting can contribute to the development of a more secure, sustainable, and user-aware IoT ecosystem.

## CCS CONCEPTS
• **Security and privacy** → **Threat hunting/detection**.

## KEYWORDS
Threat Hunting, IoT, Security, Longevity, Sustainability

## 1 INTRODUCTION

The Internet of Things (IoT) has revolutionized modern life, seamlessly integrating connected devices into our homes, workplaces, and cities [31]. From smart thermostats that adjust to our preferences to industrial sensors that monitor critical infrastructure, countless interconnected devices seamlessly integrate into our homes, workplaces, and cities [6]. These advancements provide several benefits, such as automating operations, optimising resource utilisation, and collecting the necessary information for making well-informed decisions. Nevertheless, the fast expansion of Internet of Things (IoT) devices also poses a substantial security challenge [46].

The inherent limitations of many IoT devices make them vulnerable to exploitation. Often designed to be low-cost and energy-efficient, these devices frequently lack robust security features, making them prime targets for cyberattacks. Exploited vulnerabilities can have cascading effects, compromising sensitive data, disrupting critical infrastructure operations, and even posing physical safety risks [79].

In addition to this issue, the lifespan of numerous IoT devices is quite limited. With the ever-changing landscape of technology and the unfortunate trend of planned obsolescence, there is a concerning increase in electronic waste and the associated environmental impact [15]. This raises a critical question: can we achieve a more sustainable IoT ecosystem where devices operate securely for extended periods?

This paper proposes proactive threat hunting as a potential solution to this challenge. Unlike traditional reactive methods that focus on identifying threats after they occur, proactive hunting takes a preemptive approach. By continuously monitoring for suspicious activity and potential vulnerabilities, proactive threat hunting can identify and mitigate risks before they can be exploited. This approach has the potential to significantly enhance the security posture of IoT devices, ultimately leading to extended lifespans and a more sustainable IoT landscape [23]. Additionally, a robust threat hunting strategy can contribute to the overall resilience of the IoT ecosystem. By continuously monitoring for suspicious activity and potential vulnerabilities, proactive threat hunting can identify and mitigate risks before they can be exploited. This approach has the potential to significantly enhance the security posture of IoT devices, ultimately leading to extended lifespans and a more sustainable IoT landscape [21]. This research delves deeper into this proposition by exploring the following key research questions:

(1) **Differentiating Threats in IT vs. IoT:** How does the threat landscape differ between (IT) systems and IoT? Additionally, we will explore the specific taxonomy of threats within the IoT environment.
(2) **Threat Hunting for IoT Security:** Can the principles of threat hunting be effectively applied to the unique challenges of securing IoT devices and networks?
(3) **User-awareness:** How can enhance user awareness and education regarding IoT security risks and best practices?

(4) **Environmental Impact of Secure IoT:** What are the tangible environmental benefits of proactive threat hunting in terms of extending the lifespan of IoT devices and reducing electronic waste?
(5) **Communicating Threat Hunting Insights:** How can the valuable insights derived from proactive threat hunting activities be effectively communicated to end-users to foster a more security-conscious IoT ecosystem?

By addressing these critical research questions, this study aims to provide a comprehensive framework for leveraging proactive threat hunting to achieve a more secure and sustainable future for the interconnected world of the Internet of Things.

## 2 BACKGROUND
### 2.1 IoT vs IT Attack Surface

The landscape of cybersecurity presents distinct challenges when considering attacks targeting Internet of Things (IoT) devices compared to traditional Information Technology (IT) systems [75]. One of the primary distinctions lies in the attack surface. IoT devices typically operate with constrained processing capabilities and limited resources, which often leads to a lack of effective security measures. The fundamental limitation greatly increases their vulnerability to a wide variety of cyber attacks, expanding their attack surface [32].

Another notable difference is the diversity of devices within the IoT ecosystem. As opposed to the generally uniform nature of IT equipment, IoT includes a wide range of devices that have different physical designs, platforms, and networks connectivity [52]. The presence of heterogeneity makes it difficult to adopt standardized security procedures universally, which in turn increases the vulnerability of some devices to targeted attacks and exploitation.

Furthermore, the potential physical impact of IoT attacks sets them apart from conventional IT attacks. IoT devices increasingly are integrated into critical infrastructure and life-sustaining systems, such as medical equipment and industrial control systems [35]. Consequently, successful attacks can go beyond just data breaches or service interruptions. The consequences of a successful cyberattack can extend beyond mere data breaches Instead, they can lead to severe physical consequences, posing risks to human safety and well-being.

### 2.2 The Pyramid of Pain and Prioritizing Indicators of Compromise (IOCs)

In the realm of cybersecurity, particularly in the context of IoT, understanding and addressing threats effectively is crucial. One of the frameworks that help in conceptualizing the challenge of dealing with adversaries is the "Pyramid of Pain" [36]. This model categorizes various types of indicators of compromise (IOCs) based on how much "pain" they cause to the attackers when security teams are able to detect and mitigate them [36]. The Pyramid of Pain, illustrated in Figure 1, consists of several levels, each representing a different type of IOC. From the bottom to the top, these levels are:

**Hash Values.** Hash values like MD5 and SHA1 are linked to specific malicious files, providing unique references to particular malware or files involved in an attack. Although hash values are readily accessible to security teams, they may be altered by malicious actors, rendering them an inadequate method for identifying and preventing an attack.

**Ip Addresses.** Although IP addresses are among the more prevalent IOCs, even the most experienced threat actors rarely employ their own IP addresses in an attack. VPNs and proxies can be used to change IP addresses as needed, making their denial a bypassable barrier for attackers. A defense system that relies solely on IP address restrictions to stop an attack is not an effective deterrent to threat actors, and the attack attempt will likely still succeed.

**Domain Names.** Compared to IP addresses, it is difficult to change domain names on your own. While it is more challenging for threat actors to successfully attack an organization if they are denied the use of their domains, there are ways for them to recover. Dynamic domain name system (DDNS) services and domain-generated algorithms (DGA) enable attackers to alter domain names using APIs, making it relatively easy for them to bypass domain name restrictions.

**Network/Host Artifacts.** Artifacts of an activity provide the security team with a clear distinction between malicious and benign network or host operations. URL patterns, command and control information, registry objects, files and folders, and other items may all be considered artifacts. Denying network/host artifacts causes a significant pain point for threat actors and may slow down a successful attack.

**Tools.** Threat actors are always developing the tools they utilize, resulting in developed capabilities that raise the aggressiveness of an attack. Scanning tools for vulnerabilities, creating and deploying malicious code, and executing brute force attacks to steal credentials are all widely used and are critical components of a threat actor's strategy and success. Utilizing traffic patterns or signatures to prevent threat actors from utilizing their tools is difficult for attackers to overcome.

**Tactics, Techniques, and Procedures (TTPs).** A threat actor's TTPs are a description of their approach, which includes everything from their behaviors to the particular manner in which they apply those behaviors to an attack. When an organization is able to respond to attacks on this level, it is in a position to actively defend against the practices of threat actors in general. This provides the organization with a strategic advantage that extends beyond just removing components of an attacker's metaphorical toolbox. The threat actors' chances of successfully carrying out their attack are significantly reduced when they are forced to defend at this level, which presents major challenges to them.

**Prioritizing Indicators of Compromise in IoT.** In the IoT context, prioritizing IOCs involves focusing not just on detecting them, but also on understanding the implications of each type of IOC at different layers of the IoT architecture [1]. For instance:

- **Perception and Abstraction Layers:** Focus might be more on hash values and IP addresses, as these layers deal with data collection and initial processing.
- **Network and Transport Layers:** Here, network artifacts and domain names become crucial, as data transmission and external communications are key.

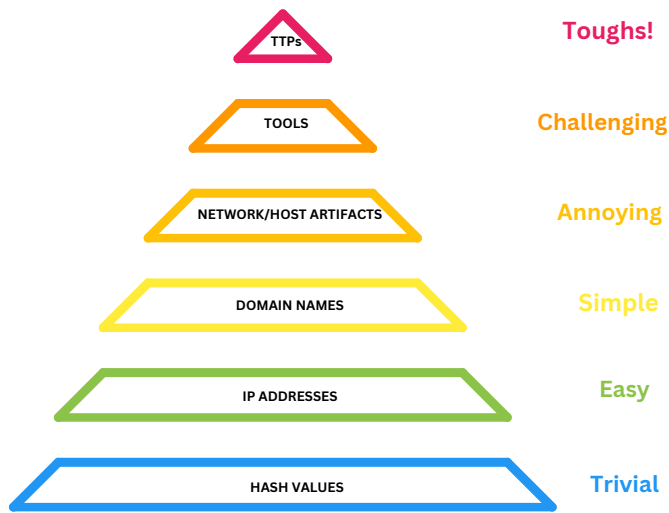

Figure 1: Pyramid of Pain

- **Application and Operation Layers:** Tools and TTPs are particularly important at these higher levels, where complex interactions and user-facing applications are managed.

By aligning the Pyramid of Pain with the layered architecture of IoT, organizations can more effectively prioritize their security efforts, focusing on the IOCs that will most disrupt potential attackers and reduce the overall risk to the system. This strategic approach not only enhances the security posture but also optimizes the allocation of resources in defending against threats [80].

## 2.3 Taxonomy of Threats in IoT

Understanding the diverse range of threats that can target different layers of IoT architecture is crucial for developing effective security measures and proactive threat hunting strategies [43]. This section provides a comprehensive taxonomy of potential threats (shown in Figure 2) across the various layers of the IoT ecosystem.

**Perception Layer.** The perception layer, also known as the sensor layer, is responsible for collecting data from the physical environment through sensors, RFID tags, and other data acquisition devices. This layer is susceptible to a range of threats, including eavesdropping, battery drainage attacks, and physical tampering, which can lead to direct damage to devices. Additionally, there are other types of attacks at this layer that aim to corrupt data, such as malicious data injection and the unauthorized disclosure of sensitive information [30].

**Abstract Layer.** The abstraction layer acts as an intermediary between the hardware (perception layer) and the software layer. It processes and abstracts the raw data collected by sensors into a more understandable format. This layer faces various threats, including node replication attacks aimed at disrupting network operations; unauthorized access that compromises system security; man-in-the-middle attacks that intercept and manipulate communications; spoofing attacks that impersonate legitimate entities; and rogue device insertion, which introduces unauthorised devices that can result in additional attacks or data breaches [47].

**Network Layer.** The network layer is responsible for transmitting the processed data from the abstraction layer to different devices, servers, or cloud services. It involves various communication protocols and networking technologies such as Wi-Fi, Bluetooth, LTE, and more. Unauthorized Access, MITM, Eavesdropping, Fragmentation Attacks, and DDoS are some of the potential threats that might occur at this layer [57].

**Transport Layer.** The transport layer manages the end-to-end communication between devices and servers, ensuring reliable data packet transfers and handling potential transmission errors. Protocols like TCP and UDP operate at this layer [73]. Threats may include:

- **Jamming Attacks:** Intentionally disrupting network services by blocking or interfering with the transmission of signals.
- **Data Injection Attacks.** Inserting incorrect or malicious data into the system, potentially leading to faulty decisions or actions.
- **Unfair Access Exploitation:** Manipulating network protocols to prioritize certain data or devices unfairly, causing service degradation or denial-of-service conditions.
- **Network Congestion Attacks:** Intentionally overloading the network with excessive traffic to degrade service quality or availability.
- **Hello Flood Attacks:** Repeatedly sending introductory packets to overwhelm the system and consume resources.

**Computing Layer.** The computing layer, also known as the processing layer, handles the storage, analysis, and processing of data received from the transport layer. Advanced data processing algorithms, including machine learning and artificial intelligence, are employed at this layer. Potential threats include malicious attacks, SQL injection attacks, data integrity breaches, virtualization exploits, unauthorized software modifications, and identity theft, all posing risks to device integrity and security [13].

**Operation Layer.** The operation layer is where the management and administration of IoT devices and their interactions occur. It includes the provisioning, configuration, and maintenance of devices, as well as the orchestration of applications and services running on the IoT platform. This layer ensures the smooth operation of the IoT system, managing resources, and optimizing performance and security. Fake Information, Bad-Mouthing attacks, Unauthorized Access, Stealing User Critical Information are some examples of threats at this layer [71].

**Application Layer.** The topmost layer of the IoT architecture, the application layer, is where the actual user applications reside. This layer translates the processed data into actionable insights and user-friendly applications, providing tailored services to end-users. Applications can range from smart home automation and industrial monitoring to health tracking and smart city management, depending on the specific needs and objectives of the IoT deployment [49]. Threats at this level include but not limited to:

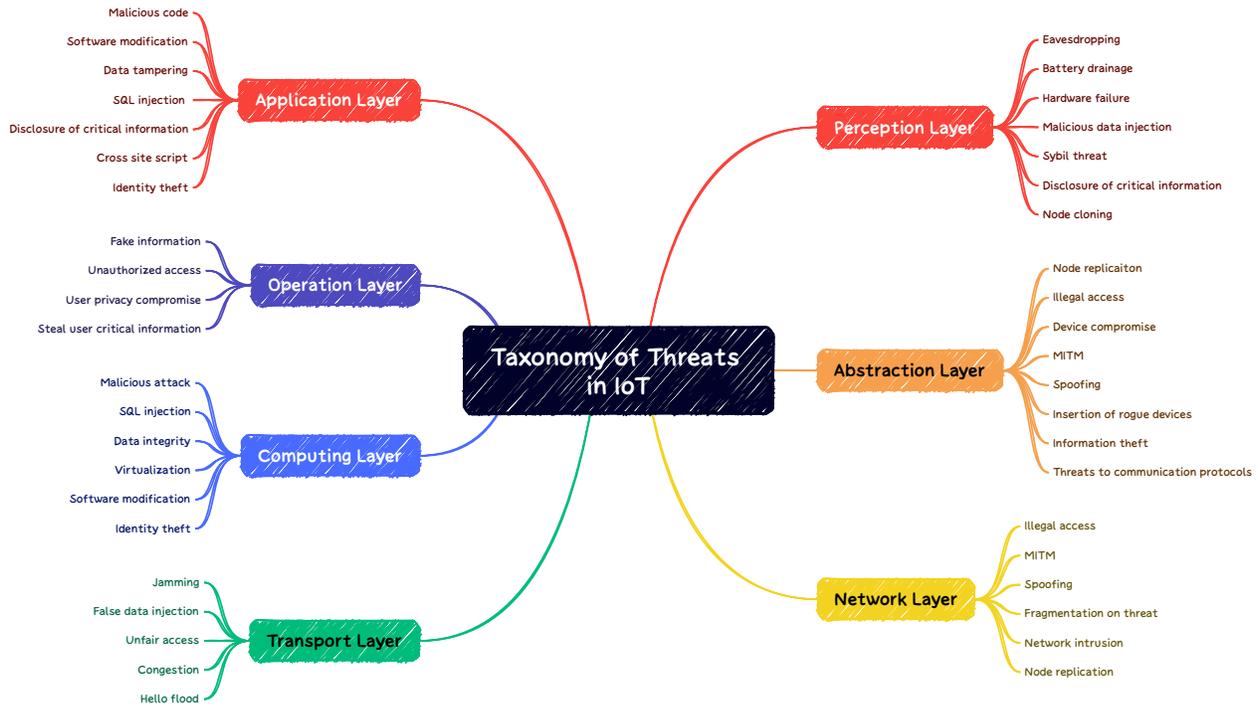

Figure 2: Taxonomy of threats in IoT

- **Malicious Code:** Software designed to harm or exploit systems.
- **Software Modification:** Altering software to perform unintended or harmful actions.
- **Data Tampering:** Manipulating data to benefit the attacker or harm the data integrity.
- **Disclosure of Critical Information:** Revealing sensitive data intentionally or accidentally.
- **Cross-Site Scripting (XSS):** Injecting malicious scripts into web pages viewed by other users.

Understanding the diverse range of threats across the various layers of the IoT architecture, contributes to development of more targeted and effective proactive threat hunting strategies. Having a deep understanding of the threats' characteristics emphasises the importance of taking a multi-layered approach for IoT security and addressing vulnerabilities and potential attack vectors at every level of the architecture.

## 2.4 Threat Detection vs. Threat Hunting vs. Threat Intelligence

**Threat Detection.** This approach operates in a reactive manner, focusing on identifying known hacking techniques and attack vectors in real-time event streams. It relies on pre-defined rules and conditions that trigger alerts upon encountering suspicious activity. These alerts are then investigated by security personnel or automated response systems. While threat detection is effective in identifying common threats, it can be susceptible to evasion tactics employed by sophisticated attackers. Additionally, the reliance on predefined rules might miss novel or zero-day threats that have not been incorporated into the detection logic [40].

**Threat Hunting.** This proactive approach assumes that adversaries may have already infiltrated the network. Threat hunters actively search for traces of past and present attacks within the event streams of connected devices. Unlike threat detection, which focuses on alerts, threat hunting involves security experts formulating hypotheses based on suspicious activity observed in the data. These hypotheses leverage creative and flexible methodologies, often drawing upon global threat intelligence feeds and historical logs from a wide range of endpoints. This proactive approach allows threat hunters to uncover hidden threats that might bypass traditional detection systems [67].

**Threat Intelligence.** Threat intelligence acts as the fuel for both threat detection and threat hunting. It involves ongoing process of gathering, analyzing, and sharing information on potential threats, vulnerabilities, and attacker tactics, techniques, and procedures (TTPs). This intelligence is derived from various sources, including internal security logs, industry reports, vulnerability databases, and open-source intelligence (OSINT). By continuously enhancing threat intelligence, companies are enabled to customize their detection and hunting strategies and effectively deal with crucial threats that occur in their environment [76].

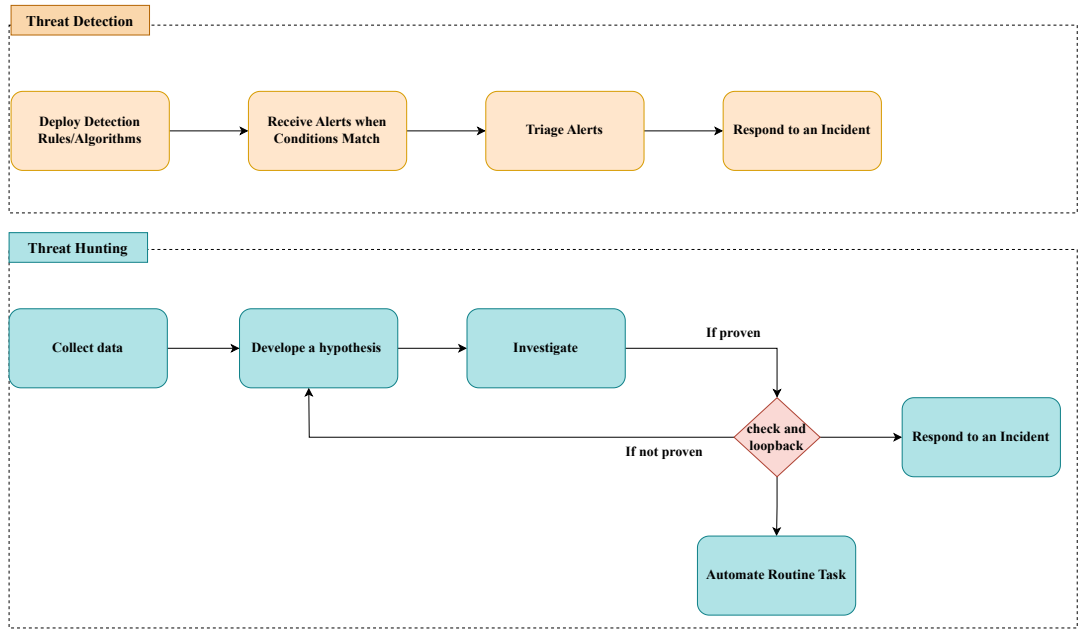

Figure 3: Threat Detetion vs. Threat Hunting

## 2.5 Security Teams and their Roles in IoT Threat Management

The effective implementation of threat detection, threat hunting, and threat intelligence requires a collaborative effort from various security teams within an organization. Understanding the distinct roles and responsibilities of these teams is crucial for building a robust security posture for IoT deployments [62].

**Security Operations Center (SOC).** The SOC serves as the central hub for real-time security monitoring and incident response. They leverage threat detection solutions to analyze network traffic, system logs, and IoT device activity and identify anomalies and suspicious behavior that align with predefined rules and indicators of compromise (IOCs). Upon detection, SOC analysts investigate alerts, assess their severity, and initiate remediation procedures. They work closely with threat intelligence teams to update detection rules based on the latest threat information [61].

**Red Team.** Red teams act as simulated attackers, employing penetration testing methodologies to identify vulnerabilities in the defenses of an organization. This also includes assessing vulnerabilities within the IoT ecosystem. Analysts at Red team actively try to exploit weaknesses in devices, communication protocols, and configurations. By mimicking real-world attacker tactics, red teams help identify blind spots in existing threat detection and hunting strategies. The findings of the red team exercises provide valuable insights for strengthening the overall security posture of IoT deployments [72].

**Blue Team.** Blue teams are responsible for defending the organization's network infrastructure, including its IoT devices. They leverage the information gleaned from threat detection, threat hunting, and red team activities to implement security controls and strengthen network configurations. Additionally, blue teams work with SOC analysts to refine the detection rules and response procedures for identified vulnerabilities [65].

**Threat Intelligence Team.** The threat intelligence team gathers, analyzes, and disseminates information on potential threats targeting IoT environments. In order to create a comprehensive picture of the evolving threat landscape, threat intelligence team have to leverage various sources, including internal logs, industry reports, vulnerability databases, and open source intelligence (OSINT). This intelligence is then used to inform and enhance the effectiveness of both threat detection and threat hunting efforts [3].

The successful implementation of these security strategies requires strong collaboration between different teams. Sharing threat intelligence across SOC, red teams, blue teams, and threat intelligence teams fosters an in-depth approach to security. By combining real-time threat detection with proactive threat hunting and a robust threat intelligence program, organizations can significantly improve their ability to identify and respond to emerging threats within their IoT deployments [20].

## 2.6 Complementary Approaches to IoT Security

While threat detection and threat hunting are crucial pillars of IoT security, other concepts play a vital role:

**Vulnerability Management.** This involves identifying, prioritizing, and improving security weaknesses within IoT devices. By proactively patching vulnerabilities, organizations can significantly reduce the attack surface and make exploitation attempts less successful [56].

**Security Hardening.** This process involves configuring IoT devices with the most secure settings possible, often by disabling unnecessary features and services. By minimizing the attack surface, security hardening can make exploitation attempts more difficult [16].

**User Education.** Raising awareness among users about the importance of IoT security is vital. Educating users on best practices for securing their devices, such as using strong passwords and keeping firmware updated, can significantly enhance overall security posture [53].

Understanding these different approaches and their limitations is crucial for developing a comprehensive IoT security strategy. By combining proactive threat hunting with robust vulnerability management, security hardening, and user education, organizations can create a more sustainable and secure IoT ecosystem.

## 2.7 Traditional Threat Hunting

Traditional threat hunting methodologies are developed with enterprise networks in mind, where there is typically a robust security infrastructure. These methodologies involve proactive and iterative searches through networks to detect and isolate advanced threats that evade existing security solutions. Key components often include sophisticated data analytics tools and extensive human expertise, that leverage both real-time and historical data to identify anomalies [58].

**Challenges of Applying Traditional Methods to IoT.**

- **Resource Constraints:** IoT devices are often designed to be cost-effective and energy-efficient, which usually means they have limited processing power, memory, and storage. This starkly contrasts with the resources available in traditional enterprise environments. Complex computations, extensive data logging, and real-time analysis required for effective threat hunting can exceed the capabilities of many IoT devices [70].
- **Scalability Issues:** IoT ecosystems typically comprise a vast number of devices. Estimates suggest that by 2025, there could be more than 75 billion IoT devices worldwide [6]. Traditional threat hunting workflows, which are often semi-manual and require significant human oversight, are not designed to handle such scale. Effectively automating these procedures while maintaining the integrity and level of detail in threat hunting is a significant challenge.
- **Diversity and Heterogeneity:** IoT devices vary widely in their operating systems, hardware configurations and functionality. This diversity complicates the implementation of uniform security measures and threat hunting techniques that are typically designed for more homogeneous enterprise environments. Adapting traditional methods to work effectively across such a varied landscape requires extensive customization, increasing complexity and cost [18].
- **Connectivity and Network Issues:** Unlike traditional networks, IoT devices may connect via a range of protocols (e.g., Wi-Fi, Bluetooth, Zigbee) that might not offer robust security features. Additionally, these devices often operate on networks that are continuously changing, with devices frequently joining and leaving. Maintaining a consistent and secure monitoring environment under these conditions is challenging [24].
- **Limited Endpoint Protection:** When used in traditional scenarios, endpoint protection systems have the ability to collect an extensive amount of information about potential threats. On the other hand, owing to their limited resources, many Internet of Things devices are unable to offer these solutions. It is difficult to effectively identify and respond to threats when there is a lack of visibility at the endpoint [38].
- **Privacy and Data Sensitivity:** IoT devices often collect sensitive data, which can include personal information. Ensuring the privacy and security of this data while conducting threat hunting activities adds another layer of complexity, particularly under stringent regulatory frameworks like GDPR [14].

To address these challenges, it is crucial to develop lightweight, scalable, and adaptable threat hunting tools specifically designed for IoT environments. This might include the use of edge computing to process data locally on devices and reduce the need for extensive data transmission and processing power [26]. Additionally, employing machine learning models that can adapt to the diverse and dynamic nature of IoT ecosystems can help in scaling threat detection mechanisms without overwhelming the network or human analysts [19].

## 2.8 The Necessity of IoT Threat Hunting

Given the unique challenges of the IoT landscape, traditional threat hunting methodologies need to be adapted to be more effective. A tailored approach should consider the following:

**Lightweight Data Collection and Analytics.** Since resource-constrained devices cannot accommodate complex data analytics, threat hunting methods for IoT should rely on lightweight data collection techniques that minimize the processing burden on the devices. Extracted data should then be forwarded to a centralized analytics platform where it can be correlated and analyzed for suspicious activity [55].

**Leveraging Machine Learning and Anomaly Detection.** Machine learning algorithms can be powerful tools for threat hunting in IoT environments. These algorithms can be trained on historical data to identify patterns of normal device behavior and then flag deviations that might indicate potential threats. This approach can help to automate threat detection and reduce the need for manual intervention [2].

**Integration with Threat Intelligence Feeds.** As with traditional threat hunting, incorporating threat intelligence feeds into the IoT threat hunting process is crucial. These feeds provide valuable insights into the latest threats and attacker tactics, allowing security teams to prioritize their hunting efforts and focus on the most relevant threats to their specific IoT ecosystem [77].

**Collaboration and Information Sharing.** The ever-evolving threat landscape necessitates collaboration and information sharing among security professionals and organizations. By sharing knowledge about IoT vulnerabilities and attack techniques, the security community can develop more effective threat hunting strategies and improve the overall security posture of the IoT landscape [11].

By adopting a tailored approach that incorporates these elements, security teams can overcome the challenges of traditional threat hunting in IoT environments and proactively identify and mitigate threats before they can cause significant damage.

## 3 PROACTIVE THREAT HUNTING

The ever-expanding landscape of the Internet of Things (IoT) presents a unique challenge: balancing the security of resource-constrained devices with environmental sustainability. Traditional reactive security approaches, where threats are identified only after they occur, often lead to compromised devices and premature replacements due to security vulnerabilities [5]. This section explores proactive threat hunting as a potential solution to this growing concern.

### 3.1 Conceptual Framework

Proactive threat hunting is a security methodology that actively searches for indicators of compromise (IOCs) and potential threats within an environment before they can be exploited [66]. In contrast to proactive threat detection, which identifies known attack patterns an attack has occurred, proactive hunting assumes the presence of adversaries within the system. It utilizes threat intelligence feeds, historical data analysis, and anomaly detection techniques to identify suspicious activity that might evade traditional detection methods [4].

*Think of it this way:* Proactive threat hunting is like a security detective, constantly searching for clues and patterns that might indicate a potential attack. It proactively investigates suspicious activity before it can escalate into a ultimate security breach.

### 3.2 Benefits for Sustainability

Proactive threat hunting offers several advantages that contribute to a more sustainable IoT ecosystem:

**Early Threat Detection and Mitigation.** By identifying vulnerabilities and potential threats before they are exploited, proactive hunting reduces the risk of successful cyberattacks and device compromise. Secure devices are less likely to require premature replacement due to security breaches, thereby extending their lifespan and reducing the environmental burden associated with electronic waste (e-waste) [17].

**Improved Operational Efficiency.** Secure devices are less susceptible to malfunctions caused by cyberattacks. Reliable and efficient operation translates to a longer device lifespan, contributing to a more sustainable IoT landscape. Imagine a smart thermostat that is constantly under attack – not only is the security of your home compromised, but the device itself might malfunction due to the attack, necessitating replacement and creating additional e-waste [41]. By proactively addressing threats, threat hunting helps maintain the smooth operation of these devices, promoting a more sustainable IoT ecosystem in the long run.

**Reduced Resource Consumption.** Extending device lifespans through proactive threat hunting can potentially reduce the resource consumption associated with manufacturing and disposing of new devices. This includes the extraction of raw materials, energy used in production and transportation, and the environmental costs of e-waste disposal. By keeping devices secure and operational for longer, we can minimize the environmental impact of the ever-growing IoT landscape [25].

**Resilience and Continuity.** By proactively identifying and mitigating threats, organizations can maintain the resilience and continuity of their IoT operations, reducing the risk of cascading failures and ensuring the reliable delivery of critical services [7].

### 3.3 Challenges and Considerations

While proactive threat hunting offers significant benefits for sustainable IoT, it's not without its challenges:

**Resource Constraints.** Traditional threat hunting methods can be resource-intensive, requiring significant processing power and memory. This poses a challenge for resource-constrained IoT devices. Developing lightweight approaches or adaptations of existing methods specifically tailored for the limitations of these devices is crucial for successful implementation [29].

**Privacy Concerns.** Data from IoT devices may be collected and analyzed as part of proactive threat hunting. It is critical to resolve privacy issues by using data anonymization methods and following data privacy standards. Finding a balance between effective security and user privacy is critical.

**False Positives.** Threat hunting can generate false positives, leading to unnecessary investigation and resource expenditure. Strategies for minimizing false positives and streamlining the investigation process need to be addressed to ensure the efficiency of proactive hunting within the IoT environment.

**Scalability.** As IoT deployments continue to grow in scale and complexity, developing scalable threat hunting methodologies that can effectively analyze data from numerous devices becomes increasingly challenging.

**Skills and Expertise.** Proactive threat hunting requires specialized skills and expertise in security analysis, data mining, and threat intelligence. Building and maintaining a skilled workforce capable of effectively conducting threat hunting activities is crucial.

Despite these challenges, the potential benefits of proactive threat hunting for sustainable IoT environments make it a compelling area for further research and development [68]. By addressing these challenges and developing tailored threat hunting methodologies specifically designed for the unique requirements of IoT ecosystems, organizations can enhance their security posture while promoting environmental sustainability.

## 4 RELATED WORK

The ever-growing realm of the Internet of Things (IoT) necessitates a delicate balance between robust security and environmental sustainability. Traditional reactive security measures, where threats are identified only after successful exploitation, often lead to compromised devices and premature replacements due to security vulnerabilities [60]. Proactive threat hunting emerges as a potential solution, offering a security methodology that actively searches for indicators of compromise (IOCs) and potential threats before they can be exploited [42]. This section delves into existing research on proactive threat hunting methodologies, with a particular focus on how these approaches can enhance security while contributing to a more sustainable IoT ecosystem by extending device lifespans and reducing electronic waste (e-waste) generation.

Table 1: Comparison of Proactive Threat Hunting Approaches for IoT Security

| Year | IoT Network Model | Dataset | ML Model | Pros (+) | Cons (-) |
|---|---|---|---|---|---|
| 2020 [81] | Edge Historical Data | CIC-DDOS2019, KDD CUP 1999 | GAN | Blockchain for Integrity | Limited Testing (Datasets) |
| 2021 [54] | Unlabeled Text | CVE Details | BERT | Generates Fake CTI | Limited Scalability |
| 2021 [78] | Edge Historical Data | CIC-DDOS2019 | CNN | Improves Intrusion Detection | High Resource Consumption |
| 2021 [22] | CTI Collection | WebText | FNN | Generates Fake CTI (Transformers) | No Detection for Poisoning |
| 2022 [69] | IoT Devices | UNSW-NB15, NSL-KDD | CNN | Efficient FL Models | Achieved High F1-Score |
| 2022 [37] | Separate Architecture | Threat Articles | BERT-BILSTM | High Accuracy | Vulnerable to Adversarial Attacks |
| 2023 [27] | Global Model | MNIST, etc. | CNN | Efficient (against poisoning) | High Resource Requirements |
| 2023 [28] | IoT Sensors | Bot-IoT | MLP | High Accuracy | Limited Effectiveness (Zero-Day Threats) |

Table 1 presents a comprehensive comparison of proactive threat hunting approaches in IoT security across various dimensions such as IoT network model, dataset, machine learning (ML) model, and associated pros and cons.

## 4.1 AI-based Proactive Threat Hunting

While AI-based threat hunting offers powerful capabilities for anomaly detection and threat pattern recognition, directly deploying complex AI models on resource-constrained IoT devices can be impractical due to computational limitations. Researchers are actively exploring techniques that enable AI-powered threat hunting while maintaining device efficiency within sustainable IoT environments [66].

**Federated Learning (FL).** FL is a promising approach that enables multiple devices to collaboratively train a machine learning model without directly sharing their data. Each device trains a local model on its own data and then shares the model updates with a central server. The central server aggregates these updates to create a global model that is then distributed back to the devices for further training. This iterative process allows the devices to collectively learn from each other's data without compromising privacy [74].

FL's distributed training approach reduces the computational burden on individual devices, preserving battery life and processing power. This can significantly extend device lifespans, as less resource-intensive operation reduces wear and tear. Additionally, FL eliminates the need for large, centralized training datasets, which can be energy-intensive to store and manage. This translates to a more environmentally friendly approach to threat hunting within large-scale IoT deployments [74].

Consider a network of smart light bulbs deployed in a large office building. Through FL, these bulbs can collaboratively train a model to identify unusual lighting patterns or energy consumption spikes that might indicate malfunctioning hardware or potential cyberattacks. For instance, the model might detect a bulb that remains unusually dim or bright for extended periods, potentially signifying a hardware issue or an attempt by a hacker to disrupt the building's lighting system. By detecting these threats early on, security professionals can address them before they cause permanent damage to the bulbs, thus extending their lifespan and reducing the need for premature replacements. This not only benefits the building's energy efficiency but also minimizes the environmental impact associated with e-waste generation [48].

**Knowledge Distillation.** This technique involves training a complex, high-performance model on a large dataset and then compressing it into a smaller, more lightweight model through a process of transferring knowledge from the teacher (complex model) to the student (lightweight model). The smaller model can then be deployed on resource-constrained devices while still retaining a significant portion of the accuracy of the original model. Knowledge distillation offers a trade-off between accuracy and resource consumption. By deploying a smaller, more efficient model on IoT devices, knowledge distillation reduces the computational demands placed on the devices, contributing to extended lifespans and reduced energy consumption. This approach enables the benefits of AI-powered threat hunting to be extended to a wider range of resource-constrained devices within the IoT ecosystem [51].

Imagine a scenario where researchers develop a powerful AI model trained on a vast collection of data encompassing known IoT threats and vulnerabilities. This data could include network traffic patterns, device logs, and exploit signatures. The complex model can then be "distilled" into a lightweight version specifically designed for deployment on smartwatches. This lightweight model can continuously analyze sensor data (e.g., heart rate, GPS location, activity levels) and network traffic patterns to detect anomalies that might indicate suspicious activity. For instance, the model might detect unusual heart rate spikes coupled with GPS data indicating the user is in an unfamiliar location, potentially suggesting unauthorized access to the smartwatch or a malware infection that manipulates sensor data. The efficient operation of this model minimizes the impact on smartwatch battery life and processing power, promoting device longevity and sustainability [45].

Here are some additional considerations for incorporating AI-based methods into sustainable IoT threat hunting:

- **Efficient Model Architectures:** Researchers are actively developing AI models specifically designed for resource-constrained devices. These models prioritize efficiency by utilizing techniques like pruning, quantization, and knowledge distillation to reduce model size and computational complexity [31].
- **On-Device Learning:** While federated learning offers a collaborative approach, advancements in on-device learning techniques are enabling devices to train lightweight models directly on their own data. This can further reduce reliance on centralized servers and communication overhead, contributing to a more sustainable approach [34].
- **Explainable AI (XAI):** As AI models become more complex, ensuring interpretability and explainability is crucial. In the context of threat hunting, XAI techniques can help security analysts understand the rationale behind the model's decisions, enabling them to prioritize investigations and validate threat detections [64].

## 4.2 Non-AI Based Proactive Threat Hunting

While AI-powered methods offer undeniable advantages in threat detection, their resource demands can pose challenges for resource-constrained IoT devices. Non-AI based proactive threat hunting approaches offer a compelling alternative for sustainable IoT security. These approaches prioritize efficiency and leverage techniques that minimize the computational burden on devices, ultimately contributing to extended lifespans and reduced e-waste generation. This section explores several prominent non-AI based methods and their potential impact on a more secure and sustainable IoT landscape.

**Lightweight Threat Hunting Techniques.**

- **Signature-based Anomaly Detection:** This approach utilizes pre-defined signatures or patterns associated with known threats to identify malicious activity. Statistical analysis of device logs or network traffic can reveal deviations from these signatures, potentially indicating an attack [10]. While lightweight and efficient, this method's effectiveness hinges on the comprehensiveness of the signature database, which needs to be constantly updated to capture evolving threats.
- **Behavioral Profiling:** By analyzing historical data on device behavior (e.g., power consumption, network traffic patterns), a baseline for normal operation can be established. Significant deviations from this baseline can trigger alerts for further investigation [12]. The accuracy of this approach depends on the availability of historical data, particularly for newly deployed devices, and might be susceptible to false positives due to variations in normal device behavior.

**Collaborative Threat Hunting.**

- **Distributed Anomaly Detection:** Leveraging data from multiple devices within a network for collective analysis offers a broader view of potential threats. This approach can identify vulnerabilities that might not be apparent from a single device perspective [33]. However, establishing secure communication channels for data sharing between devices is crucial for successful implementation. Additionally, scalability for very large deployments needs further evaluation.
- **Reputation Scoring Systems:** Assigning risk scores to devices based on their behavior and observed anomalies can help prioritize investigations and identify compromised devices within the network [59]. This approach requires careful design to ensure fairness and avoid penalizing devices exhibiting legitimate but unusual behavior.

Collaborative threat hunting offers a compelling approach for securing resource-constrained devices within the IoT ecosystem. Distributed anomaly detection and reputation scoring systems leverage network intelligence to enhance threat detection capabilities while minimizing resource consumption on individual devices. By promoting efficient device operation and facilitating early identification of security issues, collaborative threat hunting contributes to a more sustainable and secure future for the Internet of Things [44].

## 5 LONGEVITY AND SUSTAINABILITY PERSPECTIVE

**Enhancing User Awareness.** Proactive threat hunting, especially when conducted by third-party security experts, plays a pivotal role in elevating user awareness about the security of IoT devices. By implementing periodic threat assessments and introducing these assessments as universally recognized security metrics similar to the Common Vulnerability Scoring System (CVSS), these activities can standardize the security evaluation of IoT devices. This standardization allows for the inclusion of security ratings on consumer platforms such as Amazon, where potential buyers can view and compare the security robustness of IoT devices before making a purchase [9].

Further, this transparency empowers users to become active participants in the security landscape. As consumers demand a higher standard of security from device manufacturers, a cultural shift towards "security-first" design principles emerges. This incentives vendors to prioritize security throughout the development lifecycle of their products, leading to a more secure environment for all [50].

**Vendor Competition.** The introduction of security metrics as a characteristic of IoT devices serves as a catalyst for manufacturers to prioritize security enhancements. Given that security would develop a direct impact on consumer preferences and sales performance, manufacturers would become motivated to develop more secure products [63].

This competitive drive not only enhances the vendors' competence but also contributes to the overall sustainability of the IoT ecosystem. Secure devices are less likely to be compromised, reducing the frequency of replacements due to security failures and thus promoting environmental sustainability. Moreover, devices designed with robust security measures require less frequent updates and maintenance, further extending their operational lifespan and reducing electronic waste [25].

**Regulations.** The integration of security metrics into IoT device evaluation creates an opportunity for government intervention. Governments can establish regulations that mandate the inclusion of standardized security ratings on consumer products. This approach utilises market dynamics to enhance incentives for using secure development practices and fosters a fair competitive environment among manufacturers. As a result, customers are guaranteed a baseline level of security.

Furthermore, governments can enact policies that incentives the development of secure IoT devices. This could involve tax breaks for manufacturers who prioritize security features or subsidies for consumers who purchase devices with demonstrably strong security ratings. These measures encourage innovation and create a market environment that prioritizes security by design [39].

All in all, the sustainability benefits of a security-conscious IoT landscape are multifaceted. Secure devices are less susceptible to compromise, reducing the need for premature replacements due to security breaches. This not only reduces costs for customers but also contributes to environmental sustainability by minimizing electronic waste. Additionally, well-secured devices require fewer security patches and updates throughout their lifespan, further extending their operational life and reducing the environmental impact associated with device production and disposal.

Proactive threat hunting serves as the engine driving this virtuous cycle of user awareness, secure development practices, and government regulations. By continuously identifying and addressing vulnerabilities, it fosters a culture of security within the entire IoT industry. This proactive approach ensures that the rapid advancements in IoT technology are accompanied by equally robust security measures, ultimately promoting a sustainable future for the Internet of Things. In essence, proactive threat hunting empowers users, encourages secure development practices by manufacturers, and informs government regulations, all contributing to a more secure and sustainable future for the ever-evolving world of IoT [8].

## 6 CONCLUSION

The integration of proactive threat hunting into the IoT security framework marks a significant advancement towards ensuring the longevity and reliability of IoT systems. This paper has demonstrated that through continuous monitoring and the early detection of vulnerabilities, proactive threat hunting not only mitigates risks but also enhances the overall security posture of IoT devices. By adopting security metrics such as the CVSS and making these metrics visible on consumer platforms, this approach also significantly boosts user awareness and influences purchasing behaviors, encouraging a market shift towards more secure IoT products.

Furthermore, the implementation of third-party threat hunting services has been shown to be a viable solution for maintaining rigorous security standards without imposing excessive burdens on IoT manufacturers. This not only helps in fostering a security-centric culture within the industry but also promotes the development of more robust and trustworthy devices. As IoT continues to permeate all aspects of modern life, the role of proactive threat hunting will become increasingly central in balancing technological innovation with security and sustainability.

In conclusion, proactive threat hunting is not just a security measure; it is a fundamental component that will define the future landscape of IoT. By embedding security into the fabric of IoT development and consumer awareness, it paves the way for a safer, more sustainable digital future.